\documentclass[journal]{IEEEtran}

\usepackage{graphicx}
\usepackage{bm}

\begin{document}

\title{Simulation Experiment of BCI Based on Imagined Speech EEG Decoding}

\author{Kang~Wang,~Xueqian~Wang,~and~Gang~Li
}


\maketitle

\begin{abstract}
Brain Computer Interface (BCI) can help patients of neuromuscular diseases restore parts of the movement and communication abilities that they have lost. Most of BCIs rely on mapping brain activities to device instructions, but limited number of brain activities decides the limited abilities of BCIs. To deal with the problem of limited ablility of BCI, this paper verified the feasibility of constructing BCI based on decoding imagined speech electroencephalography (EEG). As sentences decoded from EEG can have rich meanings, BCIs based on EEG decoding can achieve numerous control instructions. By combining a modified EEG feature extraction mehtod with connectionist temporal classification (CTC), this paper simulated decoding imagined speech EEG using synthetic EEG data without help of speech signal. The performance of decoding model over synthetic data to a certain extent demonstrated the feasibility of constructing BCI based on imagined speech brain signal. 
\end{abstract}

\begin{IEEEkeywords}
BCI,imagined speech,sentence,without sound
\end{IEEEkeywords}

\IEEEpeerreviewmaketitle

\section{Introduction}
\IEEEPARstart{P}{atients} with neuromuscular diseases lost parts of their communication and movement abilities. That's not convenient for their daily life. Specially, for amyotrophic lateral sclerosis patients who are in the locked-in state, they even will lose the ability of eye movement. The final thing the patients can control may just be their thoughts.

BCIs are such systems that do not rely on the peripheral neurous system and muscles, and can serve patients of neuromuscular diseases to restore communication or movement ability\cite{wolpaw2002brain}. BCIs record brain activities and map them to device instructions, so a typical BCI include signal acquisition, signal feature extraction, signal classification and maybe feedback. 

The most commonly used signal acquisition method is EEG which records brain activity on the scalp in the form of electrical signal. As EEG is in the form of multi-channel electrical signal, feature extraction methods for EEG can be divided to three classes, extracting temporal feature\cite{pfurtscheller1998separability,jrad2012identification}, frequency features\cite{polat2007classification,inuso2007brain} and spatial features\cite{wang2006common,subasi2010eeg}, according to the signal attributes they care about. Most of the extraction methods focus on only one attribute and care little about the other two.

Current BCIs mostly rely on recording several classes of brain activities like motor imagery (MI) and steady state visual evoked potential (SSVEP), then map them to specific instructions . For example, BCI based on motor imgery may work through mapping left hand motor imagery to car moving forward and mapping right hand motor imagery to car moving backward. The number of brain acitivities is countable. Though applications of BCI like controling car or controlling UAV can be diverse, ability of each application is limited.     

Variable-length sentences can have rich meanings. If we can decode the sentence from imgined speech EEG signal, incorporating natural language processing will result in BCIs with flexible functions. \cite{dasalla2009single} focused on the task of imagined vowel classification. \cite{deng2010eeg} focused on classfying imagined syllables with different rhythms. \cite{martin2016word} researched on word pair classification during imagined speech. The above studies didn't consider the sentence-level imagined speech decoding. \cite{herff2015brain} decoded sentences from electorcorticography (ECoG) during speaking, but needed the recorded sound signal to help split ECoG signal and label each ECoG segment what phones they correspond to.

The above researches are all about imagined speech decoding task, but constructing BCIs needs decoding sentences from brain signal without sound information. In this paper we modified an EEG feature extraction method\cite{lawhern2016eegnet} and based on it we introduced recurrent neural network (RNN) to capture the temporal features of EEG signal. Then we combined the feature extraction structure with CTC to avoid reliance on sound signals. To verify the feasibility of decoding EEG signal without sound signal, we applied the decoding model on synthetic imagined speech EEG signal. The decoding performance on synthetic data demostrated ability of the model decoding imagined speech EEG signal without sound information and the feasibility of constructing BCIs based on imagined speech directly.
\section{methods}
In this paper, we modified a convolutional neural network (CNN) which extracts spatial and frequency features of EEG signal\cite{lawhern2016eegnet}. We introduced RNN above it to further extract temporal features and deal with the temporal nonstationarity of EEG. The modified EEG feature extraction method can extract temporal, frequency as well as spatial features and output a feature sequence.

\begin{figure*}[!t]
\centering
\includegraphics[width=6in]{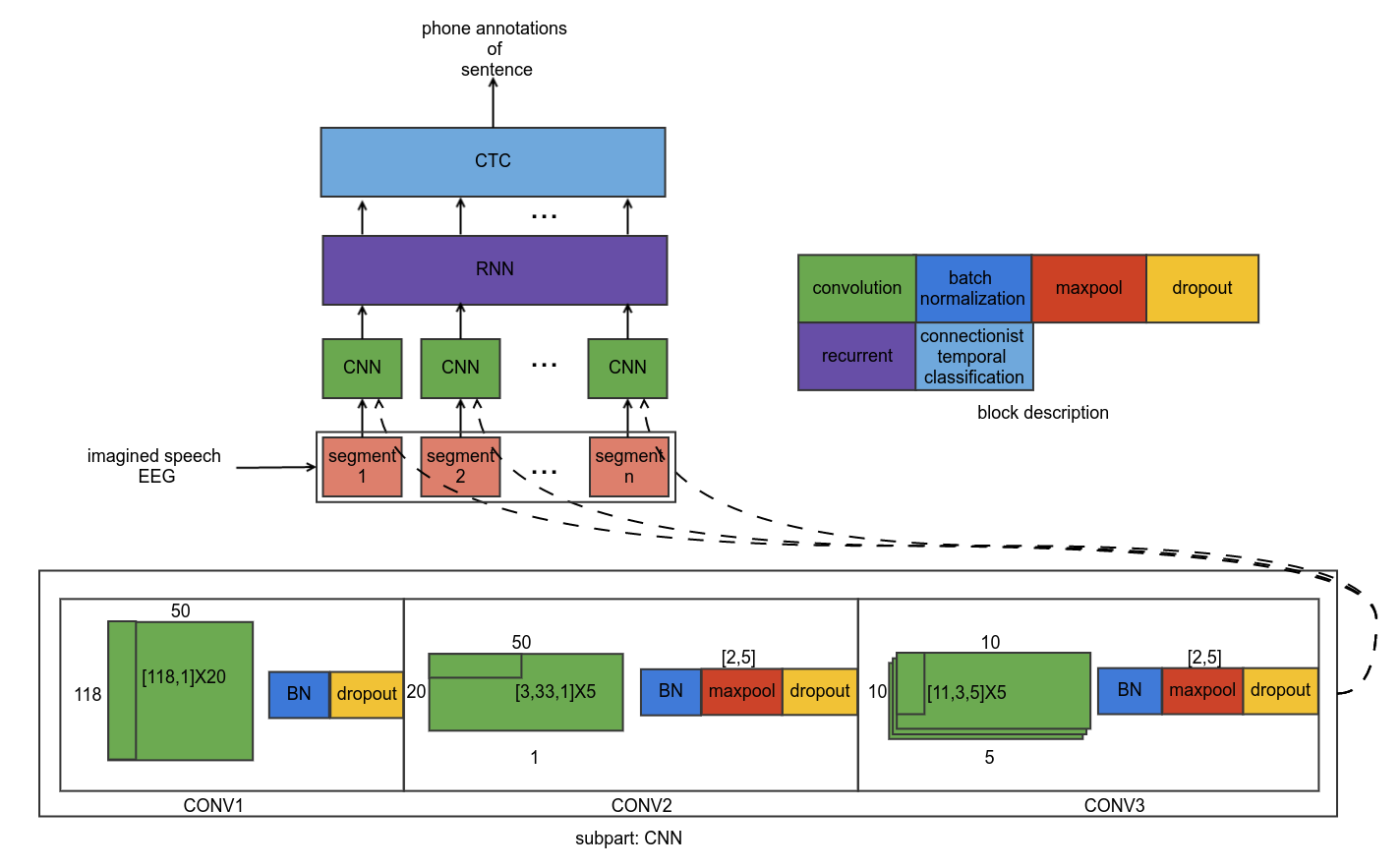}
\caption{The structure for decoding imagined speech EEG signal. Assume the signal has 118 channels and is splitted to a sequence of segments with each segment having 50 temporal samples. After extracting spatial and frequency features of each segment, RNN is applied to extract temporal features of the CNN feature sequence. The CTC layer outputs the final most possible phone sequence of the input imagined speech EEG signal.}
\label{fig:decoding}
\end{figure*}

 As we mentioned before, \cite{herff2015brain}~needed the recorded sound information to help split brain signal and assign each brain signal segment the corresponding phone. Using the labelled brain signal segment, \cite{herff2015brain}~then trained a model to classify new brain signal segments. To construct BCI based on imagined speech, no sound information is offered any more. Inspired by the work of \cite{hannun2014deep}~and \cite{assael2016lipnet}, we combined the modified feature extracture structure with CTC to avoid the reliance on sound signal. The whole structure of the decoding method is shown in Fig.\ref{fig:decoding}.

\subsection{subpart:EEGnet}
\begin{table}[!t]
\caption{EEGnet Description}
\label{tab:eegnet}
\centering
\begin{tabular}{|cccc|}
\hline
Layer&Input&Operation&Output\\
\hline
1&$118\times 50$&20 1-d Convolution($118\times 1$)&$20\times 50$\\
&$20\times 50$&BatchNorm&$20\times 50$\\
&$20\times 50$&Dropout(0.5)&$20\times 50$\\
\hline
2&$20\times 50$&Reshape&$20\times 50\times 1$\\
&$20\times 50\times 1$&5 2-d Convolution($3\times 33\times 1$)&$20\times 50\times 5$\\
&$20\times 50\times 5$&BatchNorm&$20\times 50\times 5$\\
&$20\times 50\times 5$&Maxpool(2,5)&$10\times 10\times 5$\\
&$10\times 10\times 5$&Dropout(0.5)&$10\times 10\times 5$\\
\hline
3&$10\times 10\times 5$&5 2-d Convolution($11\times 3\times 5$)&$10\times 10\times 5$\\
&$10\times 10\times 5$&BatchNorm&$10\times 10\times 5$\\
&$10\times 10\times 5$&Maxpool(2,5)&$5\times 2\times 5$\\
&$5\times 2\times 5$&Dropout(0.5)&$5\times 2\times 5$\\
\hline
\end{tabular}
\end{table}
The CNN part of the decoding model is based on \cite{lawhern2016eegnet} which proposed a compact CNN structure for EEG feature extraction called EEGnet. The EEGnet include three convolution layers. 

The first layer carries out 1-dimensional convolution to fuse the channels in original data and reduce redundancy of spatial information. The fusion of channels avoid the reliance on electrodes' spatial locations. The second layer firstly reshape the output of first layer from multi-channel temporal signal to a one-layer picture, then carries out 2-dimensional convolution on the one-layer picture with convolutional kernels whose temporal dimension is much longer than the spatial dimensiion. As temporal convolution can be considered as band filtering in the frequency domain, the second convolution layer extracts frequency features of signal. The third layer carries out 2-dimensional convolution on the multi-layer output of the second layer with convolutional kernels whose spatial dimension is longer, so the third layer extracts the spatial features of signal. 

Table \ref{tab:eegnet} gives out the description of the EEGnet. In the table we describe the EEGnet with assumption that EEG signal segment has the shape of 118 channels and 50 time samples.

\subsection{subpart:RNN}
In the decoding network, firstly the original input signal will be splitted into a sequence of segments. Each segment goes through the same EEGnet subpart for extracting frequency and spatial features. The sequence of signal segments is transformed to a sequence of CNN features. As EEG has the temporal nonstationarity, above the CNN feature sequence we use RNN to deal with the non-stationarity and further capture temporal features of the CNN feature sequence. We emplied a one-layer one-directional LSTM-cell RNN layer to achieve the modified EEG feature extraction method.

\subsection{subpart:CTC}
As we mentioned before, if the goal is constructing BCI based on imagined speech brain activity, no sound information will be offered any more otherwise speech recognition have already solved the problem. To decode a sentence from imgined speech EEG, inspired by the work of \cite{hannun2014deep}~and \cite{assael2016lipnet} whose common ground is employing a structure of combining convolutional recurrent neural network (CRNN) feature extraction with CTC, we introduced CTC to the modified EEG feature extraction method. 

As RNN can only be trained to make a series of independent label classificaitons. In the sequence labelling task, CTC is proposed to transform RNN's outputs into the final label sequence\cite{graves2006connectionist}.

Assume the CNN feature sequence $\bm{x}$ has the shape $({R^m})^T$ and the corresponding sequence label $\bm{l}$ has the shape $\mathit{L}^{*}$. ${R^m}^T$ means $T$ $R^m$ vectors, $\mathit{L}$ means the alphabet of labels and $*$ means variable length.

CTC needs adding a $blank$ label to the original alphabet to form $\mathit{L}^{'}=\mathit{L}\cup\{blank\}$. Represents the output sequence of RNN as $\bm{y}=\mathcal{N}(\bm{x})=({R^n})^T$, where $n=|\mathit{L}^{'}|$. $y_k^t$ means observing label $k$ at time $t$. Then a path in the output of RNN can be represented as :
\begin{equation}
p(\pi|\bm{x})=\prod^{T}_{t=1}y_{\pi_t}^t, \forall \pi\in {\mathit{L}^{'}}^T
\end{equation} 

Under the help of $\_$, i.e. $blank$ label, and a mapping function $\mathcal{B}:{L^{'}}^T\mapsto L^{\leq T}$:
\begin{equation}
\mathcal{B}(a\_ab\_)=\mathcal{B}(\_aa\_\_abb)=aab
\end{equation}
CTC can compute the final sequence label from RNN outputs:
\begin{equation}
p(\bm{l}|\bm{x})=\sum_{\pi\in \mathcal{B}^{-1}(\bm{l})}p(\pi|\bm{x}).
\end{equation}

When training the decoding model, the loss function over training dataset $\mathcal{S}$ is as follows:
\begin{equation}
\mbox{OBJ}(\mathcal{S})=-\sum_{(\bm{x},\bm{l})\in \mathcal{S}}ln\big(p(\bm{l}|\bm{x})\big)
\end{equation}

As the loss function focuses on the final sequence label and doesn't care about what label each input segment corresponds to, the incorporation of CTC avoids the reliance on sound information or segment labels.
\section{experiments}
As we can't find a dataset of sentence-level imagined speech EEG and limited to the performance of the equipment at hand, we synthesized sentence-level imagined speech EEG based on a dataset of imagined vowel EEG classification task. 
\subsection{imagined vowel dataset}
The imagined vowel dataset comes from \cite{dasalla2009single}. The dataset includes two classes of imagined vowels, i.e. /a/ and /u/, and a rest state which means imagining nothing. The purpose of the dataset is to classify between every two different classes of vowel imagination EEG signal. For three different classes of imgined vowels, there are three classficaitons, /a/-/u/, /a/-rest and /u/-rest. 

For each classification, there are 100 samples with each class 50 samples. In each classfication, the samples have been processed by the common spatial pattern (CSP) mehtod\cite{wang2006common}, so each sample has 4 channels which are related to the classification. For each class in each classification, we used only the 30 training samples of 50 samples.

To prepare for the synthesis, we processed the samples into three sets of samples. For the /a/ set, we choosed the 30 training samples from classification /a/-/u/ and 30 training samples from classification /a/-rest. We concatenated samples from the two groups in order in the channel dimension to form 30 samples of /a/ with 8 channels. The same process was applied to /u/ and rest class.

\subsection{synthetic sentence-level imgined speech data}
A sentence can be composed of a series of phone elements, so the idea is to concatenate a series of vowels to imitate a sentence. We now have three sets of imagined vowel EEG signal segments including /a/, /u/ and rest.

\begin{figure*}[!t]
\centering
\includegraphics[width=5in]{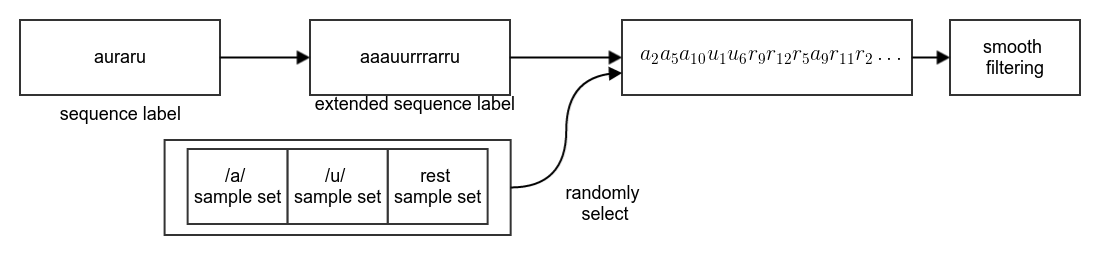}
\caption{The procedure of synthesizing sentence-level imagined peech EEG. Sequence label is generated randomly and extended randomly. According to each element in extended sequence label, a corresponding signal segment is chosen randomly. A final smooth filtering is applied to make the synthetic data more real.}
\label{fig:synthesis}
\end{figure*}

Firstly, a random sequence label is generated. As phone elements in a sentence last for different time. Secondly, each element in the generated sequnece label is extended randomly. Thirdly, to enhance the randomness of generated signal, for every element in the extended sequence label we choose from its corresponding samples set a signal segment. After concatenating all the randomly chosen signal segments in the temporal dimension and smooth filtering, a signal imitating sentence-level imagined speech EEG is generated. The procedure of the synthesis is shown in Fig.\ref{fig:synthesis}.

The synthetic EEG data and the unextended sequence labels are considered as input EEG data and corresponding sequence labels for training the decoding model.  
\subsection{scheme of model training}
Before the training, a set of test imagined speech smaples and corresponding sequence labels are generated. Before every iteration of training, 128 training samples and labels are generated. Test dataset is evaluated every 100 training iterations. 
\subsection{evaluation}
The decoding model outputs the sequence label $h(\bm{x})$ with maximum probability through greedy algorithms or dynamic programming.

The decoding performance of the model is indicated by the character-level edit distance\cite{heeringa2004measuring} over test set:
\begin{equation}
\mbox{EVAL}(\mathcal{S}^{'})=\frac{1}{|\mathcal{S}^{'}|}\sum_{(\bm{x},\bm{l})\in \mathcal{S}^{'}}\frac{\mbox{ED}(h(\bm{x}),\bm{l})}{|\bm{l}|}
\end{equation}
where $\mbox{ED}(\bm{x},\bm{y})$ means the edit distance between two sequence labels and $\mathcal{S}^{'}$ is the test dataset.
\section{results}
\subsection{training results}
After around 200 iterations of training, the decoding model converges over test dataset. Evaluation results over test dataset after 0, 100 and 200 iterations can be seen in table.\ref{tab:results}.
\begin{table}[htbp]
\caption{Evaluation Results Over Test Dataset}
\label{tab:results}
\centering
\begin{tabular}{|ccc|}
\hline
Iteration&Loss&Character-level Edit Distance\\
\hline
0&45.2&0.869\\
\hline
100&15.15&0.469\\
\hline
200&2.1&0.009\\
\hline
\end{tabular}
\end{table}

The character-level edit distance reduces to around 0 after 200 iteration. That means the trained model can almost completely give out what sequence labels the test dataset has. 
\subsection{decoding performance}
\begin{figure}[htbp]
\centering
\includegraphics[width=1.7in]{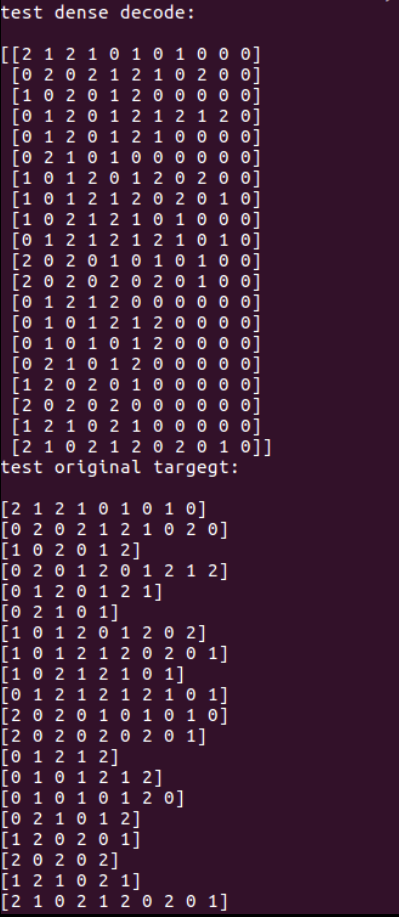}
\caption{The decoding performance over 20 test samples after 200 iterations. }
\label{fig:performance}
\end{figure}

When evaluating model on the test dataset, the first 20 decoded sequence labels and the corresponding real sequence labels are also printed for evaluating the decoding performance subjectively. The decoding performance over 20 test sampless after 200 iterations is shown in Fig.\ref{fig:performance}.

In Fig.\ref{fig:performance}, the upper part is the decoded results of trained model over 20 test samples and the lower part is the real sequence labels of the same samples. At the end of the decoded results, there are some extra 0 used to align the format.
\section{discussion}
The goal of the research is to verify the feasibiliyt of constructing BCI based on imagined speech EEG decoding. The results demonstrate that sentence-level semantic information can be decoded from EEG directly through a decoding network. If the imagined sentences can be recognized by the BCI, combining with natural language processing will result in BCI with plenty of instructions and diverse abilities.

Previous researches about imgined speech decoding have the problems of not decoding sentence-level informaiton or demanding sound information to help split and label EEG segments. The model used in this paper can decode sentence-level semantic information and avoid the reliance on sound information by incorporating CTC. 

Though the results demostrate that the decoding model can give out sequence labels of test dataset, the model was tested only on the synthetic data. Further verfications demand experiments on real imagined speech data and the data should be recorded more accurately on the Broca or Wernicke areas in the form of electrocorticography (ECoG). The combination with natural language processing should also be considered to verify the construction of useful BCI.
\section{Conclusion and future work}
In this paper, considering the problem of limited BCI instructions, we want to verify the feasibility of constructing BCI based on imagined speech brain signal decoding. We modified a EEG feature extraction structure to extract temporal, frequency and spatial features of EEG signal. Combining the modified feature extraction method with CTC, we constructed a decoding network for decoding stentence-level information and avoiding the reliance on sound information or segment labels. Experiments on synthetic imagined speech data were carried out to verify the performance of the decoding network. The results to a certain extent illustrate the feasibility of constructing BCI based on imagined speech decoding.

To further verify the feasibility, accurate real imagined speech brain signal should be recorded and verified. To construct useful BCI, the researches on combination with natural language processing should also be considered.



\ifCLASSOPTIONcaptionsoff
  \newpage
\fi

\bibliographystyle{IEEEtrans}
\bibliography{test}

\begin{thebibliography}{10}
\providecommand{\url}[1]{#1}
\csname url@samestyle\endcsname
\providecommand{\newblock}{\relax}
\providecommand{\bibinfo}[2]{#2}
\providecommand{\BIBentrySTDinterwordspacing}{\spaceskip=0pt\relax}
\providecommand{\BIBentryALTinterwordstretchfactor}{4}
\providecommand{\BIBentryALTinterwordspacing}{\spaceskip=\fontdimen2\font plus
\BIBentryALTinterwordstretchfactor\fontdimen3\font minus
  \fontdimen4\font\relax}
\providecommand{\BIBforeignlanguage}[2]{{%
\expandafter\ifx\csname l@#1\endcsname\relax
\typeout{** WARNING: IEEEtran.bst: No hyphenation pattern has been}%
\typeout{** loaded for the language `#1'. Using the pattern for}%
\typeout{** the default language instead.}%
\else
\language=\csname l@#1\endcsname
\fi
#2}}
\providecommand{\BIBdecl}{\relax}
\BIBdecl

\bibitem{wolpaw2002brain}
J.~R. Wolpaw, N.~Birbaumer, D.~J. McFarland, G.~Pfurtscheller, and T.~M.
  Vaughan, ``Brain--computer interfaces for communication and control,''
  \emph{Clinical neurophysiology}, vol. 113, no.~6, pp. 767--791, 2002.

\bibitem{pfurtscheller1998separability}
G.~Pfurtscheller, C.~Neuper, A.~Schlogl, and K.~Lugger, ``Separability of eeg
  signals recorded during right and left motor imagery using adaptive
  autoregressive parameters,'' \emph{IEEE transactions on Rehabilitation
  Engineering}, vol.~6, no.~3, pp. 316--325, 1998.

\bibitem{jrad2012identification}
N.~Jrad and M.~Congedo, ``Identification of spatial and temporal features of
  eeg,'' \emph{Neurocomputing}, vol.~90, pp. 66--71, 2012.

\bibitem{polat2007classification}
K.~Polat and S.~G{\"u}ne{\c{s}}, ``Classification of epileptiform eeg using a
  hybrid system based on decision tree classifier and fast fourier transform,''
  \emph{Applied Mathematics and Computation}, vol. 187, no.~2, pp. 1017--1026,
  2007.

\bibitem{inuso2007brain}
G.~Inuso, F.~La~Foresta, N.~Mammone, and F.~C. Morabito, ``Brain activity
  investigation by eeg processing: wavelet analysis, kurtosis and renyi's
  entropy for artifact detection,'' in \emph{Information Acquisition, 2007.
  ICIA'07. International Conference On}.\hskip 1em plus 0.5em minus 0.4em\relax
  IEEE, 2007, pp. 195--200.

\bibitem{wang2006common}
Y.~Wang, S.~Gao, and X.~Gao, ``Common spatial pattern method for channel
  selelction in motor imagery based brain-computer interface,'' in
  \emph{Engineering in Medicine and Biology Society, 2005. IEEE-EMBS 2005. 27th
  Annual International Conference of the}.\hskip 1em plus 0.5em minus
  0.4em\relax IEEE, 2006, pp. 5392--5395.

\bibitem{subasi2010eeg}
A.~Subasi and M.~I. Gursoy, ``Eeg signal classification using pca, ica, lda and
  support vector machines,'' \emph{Expert Systems with Applications}, vol.~37,
  no.~12, pp. 8659--8666, 2010.

\bibitem{dasalla2009single}
C.~S. DaSalla, H.~Kambara, M.~Sato, and Y.~Koike, ``Single-trial classification
  of vowel speech imagery using common spatial patterns,'' \emph{Neural
  networks}, vol.~22, no.~9, pp. 1334--1339, 2009.

\bibitem{deng2010eeg}
S.~Deng, R.~Srinivasan, T.~Lappas, and M.~D'Zmura, ``Eeg classification of
  imagined syllable rhythm using hilbert spectrum methods,'' \emph{Journal of
  neural engineering}, vol.~7, no.~4, p. 046006, 2010.

\bibitem{martin2016word}
S.~Martin, P.~Brunner, I.~Iturrate, J.~d.~R. Mill{\'a}n, G.~Schalk, R.~T.
  Knight, and B.~N. Pasley, ``Word pair classification during imagined speech
  using direct brain recordings,'' \emph{Scientific reports}, vol.~6, 2016.

\bibitem{herff2015brain}
C.~Herff, D.~Heger, A.~De~Pesters, D.~Telaar, P.~Brunner, G.~Schalk, and
  T.~Schultz, ``Brain-to-text: decoding spoken phrases from phone
  representations in the brain,'' \emph{Frontiers in neuroscience}, vol.~9, p.
  217, 2015.

\bibitem{lawhern2016eegnet}
V.~J. Lawhern, A.~J. Solon, N.~R. Waytowich, S.~M. Gordon, C.~P. Hung, and
  B.~J. Lance, ``Eegnet: A compact convolutional network for eeg-based
  brain-computer interfaces,'' \emph{arXiv preprint arXiv:1611.08024}, 2016.

\bibitem{hannun2014deep}
A.~Hannun, C.~Case, J.~Casper, B.~Catanzaro, G.~Diamos, E.~Elsen, R.~Prenger,
  S.~Satheesh, S.~Sengupta, A.~Coates \emph{et~al.}, ``Deep speech: Scaling up
  end-to-end speech recognition,'' \emph{arXiv preprint arXiv:1412.5567}, 2014.

\bibitem{assael2016lipnet}
Y.~M. Assael, B.~Shillingford, S.~Whiteson, and N.~de~Freitas, ``Lipnet:
  Sentence-level lipreading,'' \emph{arXiv preprint arXiv:1611.01599}, 2016.

\bibitem{graves2006connectionist}
A.~Graves, S.~Fern{\'a}ndez, F.~Gomez, and J.~Schmidhuber, ``Connectionist
  temporal classification: labelling unsegmented sequence data with recurrent
  neural networks,'' in \emph{Proceedings of the 23rd international conference
  on Machine learning}.\hskip 1em plus 0.5em minus 0.4em\relax ACM, 2006, pp.
  369--376.

\bibitem{heeringa2004measuring}
W.~J. Heeringa, ``Measuring dialect pronunciation differences using levenshtein
  distance,'' Ph.D. dissertation, Citeseer, 2004.

\end{thebibliography}

\end{document}